\documentclass[12pt]{article}
\begin{document}
\title{Plane waves and wave packets in particle oscillations}
\author{
L.B. Okun\thanks{e-mail: okun@heron.itep.ru}\hspace*{2mm}$^{\rm a}$,
M.V. Rotaev\thanks{e-mail: mrotaev@mail.ru}\hspace*{2mm}$^{\rm a,b}$,
M.G. Schepkin\thanks{e-mail: schepkin@heron.itep.ru}\hspace*{2mm}$^{\rm a}$,
I.S. Tsukerman\thanks{e-mail: zuckerma@heron.itep.ru}\hspace*{2mm}$^{\rm a}$
\\[5mm]
${\rm ^a}$ {\small\it Institute of Theoretical and Experimental Physics}\\
{\small\it 117218, B.Cheremushkinskaya 25, Moscow, Russia}\\
${\rm ^b}$ {\small\it Moscow Physics and Technology Institute}}
\date{}

%\begin{document}

\maketitle

\begin{abstract}

Manipulation with ill-defined notions creates an illusion
of a simple derivation of the standard phase of neutrino oscillations
in hep-ph/0311241 and hep-ph/0312180.
\end{abstract}

%\maketitle

Recent remark \cite{1} on a too simple (one page) derivation \cite{2}
of the standard formula for neutrino oscillations has been criticized
in \cite{3}. 

Let us remind that the derivation in \cite{2} is based on
 "the plane wave approach, in which the
massive neutrino states evolve in space and time as plane waves
$|\nu_k(x,t)> = e^{-iE_k t + ip_k x} |\nu_k>$".

The plane wave approach describes free particle.
This is confirmed by equation $E_k^2 = p_k^2 + m_k^2$ explicitly written
out on the same page in \cite{2}. As is well known, this is the standard definition
of the plane wave approach, used e.g. in the description of kaon oscillations.

Let us add that "the plane wave approach" is confronted with "the wave
packet treatment" in the first sentence on the page 3 of ref.\cite{2}.
Moreover, in the concluding paragraph of ref.\cite{3} it is stressed that the
plane wave approach presented in ref.\cite{2} "is rather attractive because
it is simple, {\it avoiding wave packet complications}, and uses a minimum
of well-motivated assumptions".

In spite of that on the first page of ref.\cite{3} it is written that in ref.\cite{1}
we did not notice that according to \cite{2} it is necessary to treat massive
neutrinos as wave packets. It was exactly our target: to separate plane waves
from wave packets. For that purpose we have considered only that part of
ref.\cite{2} where plane wave is used.
Not a single symbol representing wave packet can be found in that part of
ref.\cite{2}.

Another assumption in ref.\cite{2} is that  $t \approx x = L$, which
actually is replaced by the strict equality  $t=x=L$. This strict equality
is needed, according to \cite{3}, to get the standard phase.
In the same paragraph one can find
$t \approx x(1+{\bar{m^2}}/2E^2)$, where $\bar{m^2}$ is introduced
in ref.\cite{2}:  "$\bar{m^2}$ is the average of the squared neutrino masses".
However, "it is clear" to the author of ref.\cite{2,3} "that the correction
$x {\bar m^2} /2E^2$ to $t=x$ can be neglected".
As emphasized in ref.\cite{1}, such corrections are of the same order as
the standard oscillation phase and as such are used from time to time in the
literature to modify the standard phase by the notorious factor of 2.
On the other hand, the strict light-like equality $t=x$ is obviously
incompatible with the standard plane wave of a massive particle.

Thus ref.\cite{2} and ref.\cite{3} operate in a fanciful way with the notion
of plane wave approach and create illusion of a simple derivation of the
standard phase of neutrino oscillation.

As for the equal energy condition for clockless experiments mentioned by us
in ref.\cite{1} , we will discuss this in detail in a forthcoming paper.  \\

{\bf Acknowledgments}  \\

We are grateful to H. Lipkin and L. Stodolsky for useful comments.

\end{document}